\documentclass[aps,prl,twocolumn,10pt,showpacs,superscriptaddress]{revtex4}

\usepackage{graphicx}
\usepackage{amsmath,units}
\usepackage{mathrsfs}
\usepackage{color}

\graphicspath{{Figures/}}
\begin{document}

\title{Diffusion of a nano-wire through an obstacle field} 

\author{Dror Kasimov}

\affiliation{Raymond \& Beverly Sackler School of Physics and Astronomy, Tel Aviv
  University, Tel Aviv 6997801, Israel}

\author{Tamir Admon}

\affiliation{Raymond \& Beverly Sackler School of Chemistry, Tel Aviv
	University, Tel Aviv 6997801, Israel}

\author{Yael Roichman}
\email{roichman@tau.ac.il}

\affiliation{Raymond \& Beverly Sackler School of Chemistry, Tel Aviv
  University, Tel Aviv 6997801, Israel}

\date{\today}

\begin{abstract}
 
We report the first experimental realization of a rod diffusing in a two dimensional obstacle field following the single rod dynamics. We use a silver nanowire as our rod and two types of obstacles: repelling light beams and polymer pillars. We study the effect of hydrodynamic interactions on the transport of the rod, comparing both experimental realizations and recent simulations. We propose a new framework for analyzing the transport through such systems and predict a new superdiffusive regime of rod transport at high obstacle concentration and short times.       
\end{abstract}

\pacs{05.40.Fb	
05.40.-a, 
05.40.Jc	
05.60.Cd	
87.16.dj, 
}

\maketitle
A rod like particle moving between randomly and uniformly distributed obstacles on a plane is a toy model studied in relation to physical situations of diffusion in crowded environments. Examples for such systems include: diffusion in polymer melts \cite{Doi1978} and dense liquid crystal suspensions \cite{Lettinga2005,Lettinga2007,Allen1990}, diffusion of viruses in cell membranes \cite{Santangelo2007}, and diffusion in porous media. 
The rich and surprising dynamics emerging from the elongated nature of the diffusive particles render this model interesting also from the perspective of transport theory. For example, at low densities the center-of-mass diffusion decreases with obstacle density, as expected from Enskog theory \cite{ODell1975} for spheres. However, above a certain threshold, this trend is reversed and the diffusion coefficient increases with obstacle density. This behavior was predicted theoretically by kinetic theory \cite{Frenkel1981} and demonstrated in simulations \cite{Frenkel1983,Hofling2008,Tucker2010} assuming the rod moves { \em ballistically} between collisions with obstacles. The increase in diffusion coefficient was predicted to follow a power law of $\sqrt{n}$, where $n$ is the obstacle density. In simulations, powers between 0.3-0.8 were reported \cite{Hofling2008,Tucker2010}. 
The aforementioned results apply to an infinitely thin rod, point-like obstacles, and motion in two dimensions. If the rod thickness is finite a new confinement regime \cite{Tucker2010} appears, and if the rod is allowed to move in three dimensions the enhanced diffusion regime disappears \cite{Tucker2011}. 

The entire density dependence of the rod center-of-mass diffusion coefficient, $D_{cm}$, changes if the underlying motion of the rod between collisions is Brownian instead of ballistic \cite{franosch2009}. In this case $D_{cm}$ is constant at very low densities and decreases to a lower constant at high densities. The diffusion of a rod at high obstacle densities, both in the case of underlying ballistic motion and diffusive motion is unique for  elongated particles. 
Experimental works on this subject have been few so far, focusing on the motion of elongated objects in dense suspensions rather than through fixed obstacle fields \cite{Zheng2010,Cush2004,Lettinga2007,Lettinga2005}. Recently, a 3D study of  the movement of carbon nanotubes in porous agarose was reported focusing on the effect of rod flexibility \cite{Fakhri2010}. 

Here we present the first measurements of single rod dynamics in a static obstacle field.  We focus on the short time diffusion of rods and characterize the obstacle density effect on their transport in two different experimental realizations: one with polymer obstacles and one with virtual optical obstacles. We then apply external driving on the rods to induce ballistic-like characteristics to the otherwise Brownian motion of the rods, and finally, we introduce an analysis approach which highlights the effect of the underlying motion type (ballistic or Brownian) on the transport of rods in such systems. 

\begin{figure} [t]
	\includegraphics[scale=0.35]{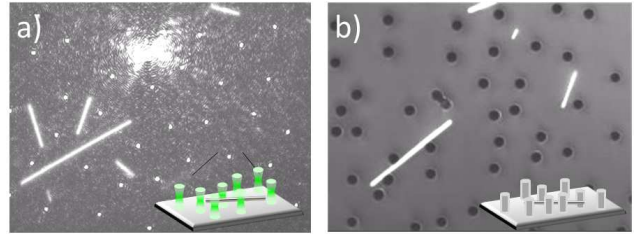}
	\caption{Two experimental realizations of a rod diffusing in an obstacle field. A silver nanowire suspended in water is subjected to a) a random field of repelling focused laser beams, and b) an array of randomly positioned polymer pillars.}
	\label{fig:experimental}
\end{figure}

Our samples consist of silver nanowires (Nanostructured and amorphous materials, AgNw, diameter $b\sim386 \pm 48$ nm, length $L\sim 5-30$~$\mu$m) dispersed in deionized water. A drop of the nanowires suspension is placed between a slide and a cover slip, both passivated with 10 \%wt solution of BSA (bovine serum albumin) to prevent nanowires from sticking to the surfaces. Our sample chamber is approximately $40 \mu m$ high, the obstacle effective height is $h\sim2 \mu m$ (for both types of obstacles \cite{supp1}). The relatively dense silver rods sediment to the bottom of the sample chamber, effectively diffusing in two dimensions. We use two methods to create randomly positioned obstacles (Fig.~\ref{fig:experimental}). The first technique is to decorate the sample floor with polymer pillars (SU8 2002 MicroChem, height and diameter $\sim 2\mu m$) using standard photolithography. The second technique uses holographic optical tweezers (HOTs) \cite{Dufresne2001,Grier2006,Polin2005,Yevnin2013} to create a random array of focused light beams. The light from the beams is reflected by the rods repelling them by momentum transfer with a force in the range of tens of fN \cite{supp1}. Additional experiments were performed with external driving of the rods. Flow was created by moving the sample (i.e. by automated motion of the microscope stage) relative to the optical scatterers array, with constant velocity.

Our HOTs are based on a Coherent Verdi laser ($\lambda=532$~nm, 6~W) and a Hamamatsu (LCOS X10468-04) liquid crystal spatial light modulator to create our soft optical scatterers \cite{Nagar2014}. Imaging takes place in an inverted microscope (Olympus IX71) in reflection mode using a CCD camera (Grasshopper3, PointGrey). A single objective (Olympus, 100x oil immersion NA=1.4) is used both for imaging and focusing of the laser light into the sample plane. The experiments in the polymer pillar configuration were performed on the same imaging system.  

A random arrangement of obstacles, at various densities, was drawn from a random distribution. A minimum distance was enforced to ensure an approximately uniform density. The obstacles' coordinates were then used both to create a photolithography mask for the polymer pillar array experiments and to create the holograms for the optical arrays. Holograms were calculated using the direct search algorithm \cite{Polin2005}. The experiments consisted of taking videos (at 10 Hz) of the movement of the rods between the different obstacle arrays. In addition, videos of freely diffusing rods were taken for reference.  
Rod position and orientation were extracted from each image by a sequential process of thresholding, identifying closed regions and extracting their properties (center of mass, orientation, and length). Consequently, a tracking algorithm \cite{crocker96} was used to identify single rod trajectories. Trajectories were usually $10-30$ min long. Only relatively long wires (L $\gtrsim5\mu m$) which retain their 2D motion were followed, while trajectories of shorter wires which tend to move in 3D were not used for calculations.
\begin{figure}[h]
\includegraphics[scale=0.17]{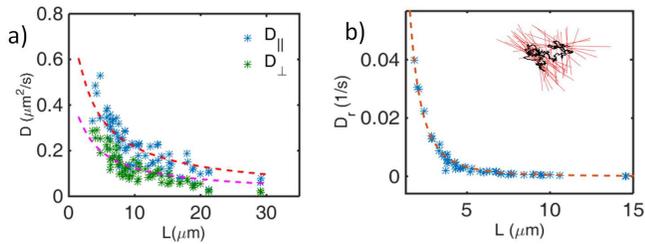}
	\caption{The diffusion coefficient of a freely moving rod as a function of rod length $L$, a) translational diffusion coefficients, $D_{\parallel}, D_{\perp}$, and b) rotational diffusion coefficient $D_r$. Fits to theory \cite{doi1986theory} with friction coeff. $\gamma=0.8$ (dashed line). Inset: A typical trajectory of a free rod.} 
	\label{fig:free_wires}
\end{figure}

We start by characterizing the diffusion of free rods in quasi 2D close to the sample floor (similarly to \cite{Duggal2006,Li2004,Colin2012}). To this end we calculate the time averaged mean squared displacement (MSD) for each rod, from which we extract the three diffusion coefficients, rotational $D_r$, longitudinal $D_\parallel$, and transverse  $D_\perp$ (Fig.~\ref{fig:free_wires}, \cite{supp2}). 

Our results agree well with theory \cite{doi1986theory}, allowing us to extract three different measurements of the effective viscosity of the solvent in the vicinity of the sample floor: $\eta^{\parallel}_s=0.984 \pm 0.055$~mPa$\cdot$s, $\eta^{\perp}_s=0.858 \pm 0.055$~mPa$\cdot$s, $ \eta^r_s=3.0 \pm 0.09$~mPa$\cdot$s. In an unbound fluid we would expect $\eta_s=0.955$ mPa$\cdot$s, which is in accord with our results for $D_{\parallel}$ and $D_{\perp}$. Surprisingly, the viscosity extracted from $D_{r}$ is much larger. A similar effect was reported previously for short actin rods diffusing in 2D \cite{Li2004}. We find the average ratio $D_{\parallel}/D_{\perp}=1.97 \pm 0.09$ in accord with the predicted value of 2. 

\begin{figure}[t]
\includegraphics[scale=0.18]{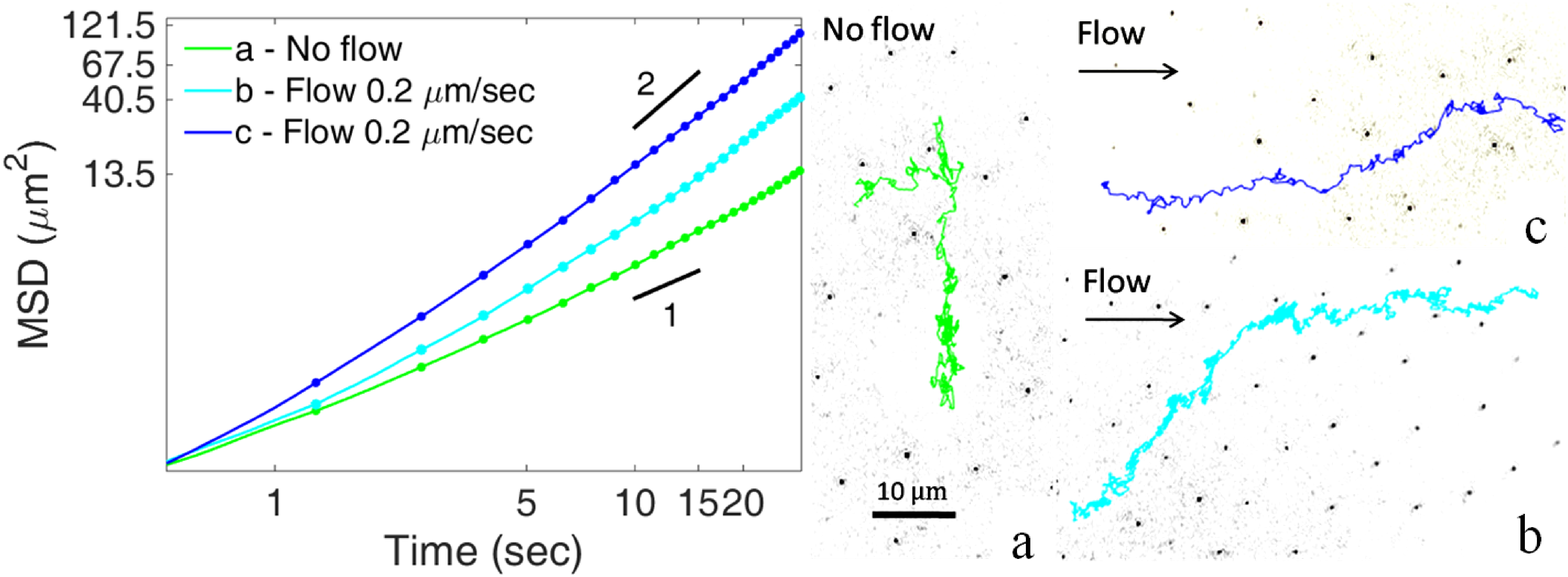}
	\caption{Log-log MSD graphs for the trajectories of nanowires 15 $\mu m$ long diffusing through an optical obstacle field array in three conditions: a) no flow, $n^* \sim 3$. b) flow - 0.2 $\mu m/sec$, $n^* \sim 5.3$. c) flow - 0.2 $\mu m/sec$,  $n^* \sim 3$. The slopes of the graphs are (from bottom to top) 1.4, 1.8 and 1.9. Right: Center of mass trajectories of nanowires at these three conditions.}
	\label{fig:flow}
\end{figure}

 
Having established the diffusion properties of the free rods, we proceed to analyze the short time diffusion of rods moving in an obstacle field (Fig.~\ref{fig:flow}), with and without external driving.

\begin{figure}[h!]
\includegraphics[scale=0.16]{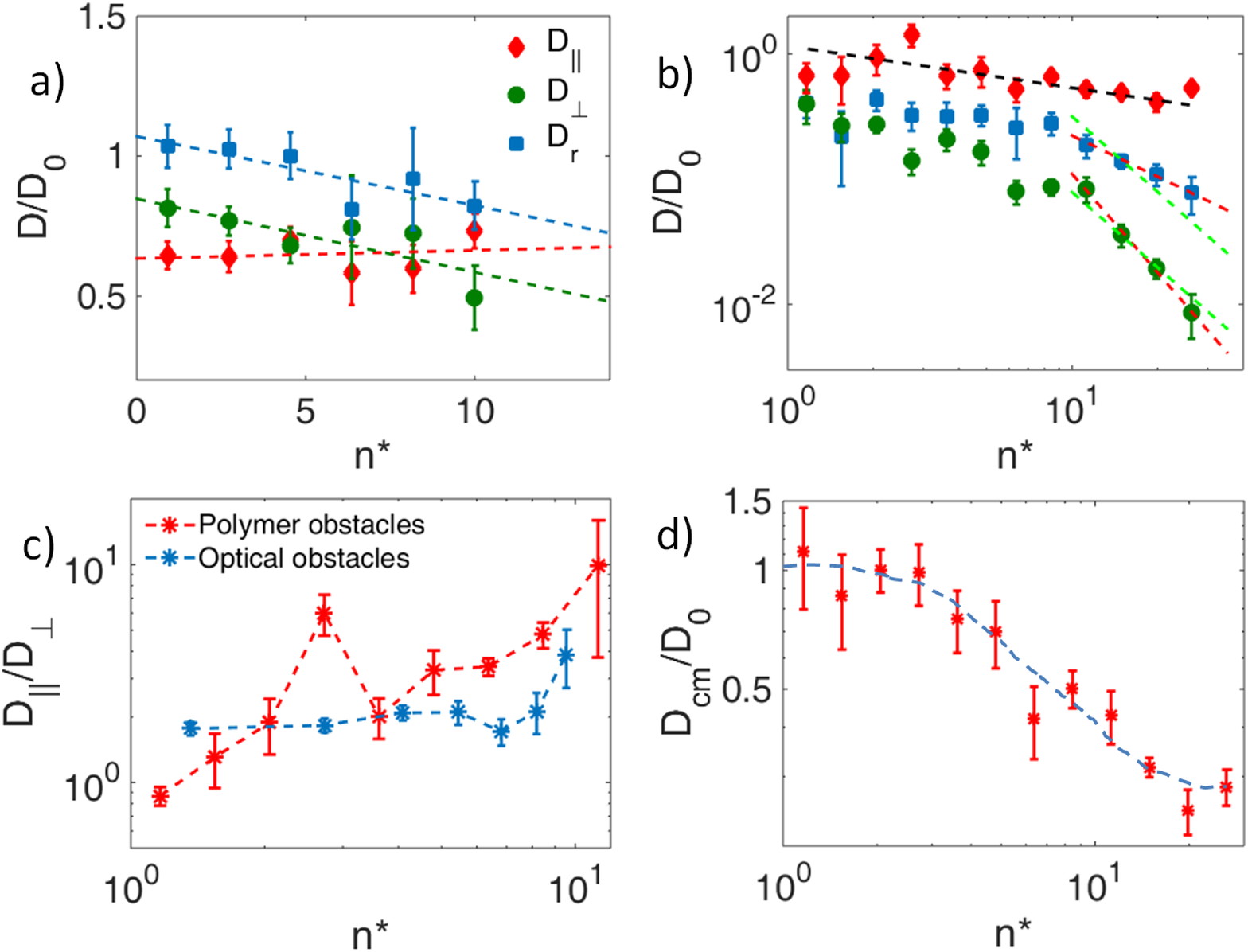}
	\caption{The normalized diffusion coefficients of a confined rod as a function of normalized density $n^*$, a) optical scatterers, and  b) polymer pillars, red dashed lines fit for $D\sim (n^*)^{-\alpha}$ at high density, green dashed line fit for $D\sim (n^*)^{-2}$.  c) Comparison of $D_\parallel/D_\perp$ for optical and polymer obstacles. d) Center-of-mass diffusion coefficient for the polymer pillar experiments.}
	\label{fig:Di}

\end{figure}

For the case of no external flow we plot the diffusion coefficients normalized by the theoretical values for free rods (Fig.~\ref{fig:Di} \cite{supp2,supp3}).  In order to compare diffusion of rods of different length and different realizations of obstacle fields, we normalized the obstacle density by the length of the rod, $n^*=n L^2=(L/\epsilon)^2$. The results from the optical scatterers setup (Fig.~\ref{fig:Di}a) show that $D_{\perp}$ and $D_r$ decrease with density while the $D_{\parallel}$ remains fairly constant, confirming previous simulation results \cite{franosch2009}.

The polymer pillar setup allows for measurements with a larger range of normalized densities up to $n^*\sim 30$ (Fig.~\ref{fig:Di}b). Here, a sharp decline after $n^*\sim 10$ is observed for both $D_{\perp}$ and $D_r$. We measure a power law decay of $\alpha\sim-2.6$ and $\alpha\sim-1.1$ for the perpendicular and rotational movement respectively. Theoretically, both declines should decay asymptotically with $\alpha\sim-2$ for infinitely thin rods. However, for rods with finite thickness it was found in simulations that $D_r\sim (n^*)^{-1}$ \cite{Cobb2005}, in accord with our measurement. Hydrodynamic interactions may also contribute to this deviation from theory. 

Although both experimental setups exhibit similar behavior, there are several differences worth highlighting. First, $D_{\parallel}$ decays slowly in the polymer pillar experiments (Fig.~\ref{fig:Di}b) contrary to its independence on obstacle density both in simulations \cite{franosch2009} and in the optical obstacles experiments (Fig.~\ref{fig:Di}a). Second, the ratio $D_{\parallel}/D_{\perp}$ increases gradually in the polymer pillar experiments, but in the optical scatterres realization it remains constant up to a threshold density over which it starts to increase (Fig.~\ref{fig:Di}c), similar to its behavior in nematic liquid crystals \cite{Lettinga2005,Allen1990}. This threshold density $n^*\sim8$ also signifies a strong change in the dependence of $D_r$ and $D\perp$ on $n^*$. The two types of obstacle arrays differ in two main ways: one, the optical obstacles do not affect flow (i.e. flow induced by rod diffusion or external driving) in the sample as opposed to the polymer pillars, and two, the optical obstacles constitute a soft repulsive potential for the wires whereas the polymer pillars constitute hard core repulsion. We believe the differences between the results of the two types of experiments arise mainly from the existence of hydrodynamic interactions (between the rod and the obstacles) in the polymer pillar experiments. One effect solid boundaries have on hydrodynamic interactions is to increase the effective viscosity of  the fluid \cite{happel}, this is due to the no-slip boundary conditions on the fluid at contact with a solid boundary \cite{Diamant09}. Here the solid obstacles between which the nanowires diffuse introduce more boundaries. As their number increases the effective viscosity increases and the diffusion coefficient decreases. This effect is manifested in the decrease of the center of mass diffusion coefficient with the increase of polymer pillars density. Additional non-trivial effects can arise from the change in the flow field due to the polymer pillars. An indication of such an effect is the difference in the ratio of the parallel to perpendicular diffusion coefficients when comparing optical to polymer obstacles. In addition, we find that the center-of-mass diffusion coefficient $D_{cm}$ in the pillar array experiments agrees only qualitatively with simulations \cite{franosch2009} and does not plateau at the expected value (Fig.~\ref{fig:Di}d).   
  
The addition of flow due to external driving of the suspending fluid relative to the obstacle field creates elongated and directional trajectories with $<R^2>~\sim~t^\alpha$ and $\alpha>1.5$ (Fig. \ref{fig:flow}b,c). It can be seen that the addition of flow excludes trajectories with back and forth motion on large scales. This observation inspired us to decouple the analysis of rod transport through obstacle fields into two characteristics: the trajectory shape, and the underlying motion along said trajectory. The shape of the path is a function of obstacle density, particle shape, and driving, while the transport type is a function of medium and driving. Back and forth motion, which occurs in diffusive transport was also not observed in simulations of ballistically moving rods \cite{Hofling2008}.

For example let us consider the effect of obstacle density on a rod scattering balistically from obstacles. Here we do not assume flow, but rather motion without Brownian dissipation, as was examined in previous simulations \cite{Frenkel1983,Hofling2008,Tucker2010}. We assume that the shape of the rod's trajectory can be described by the worm-like-chain (WLC) model for semi-flexible polymers \cite{doi1986theory}. A WLC is characterized by the extended polymer chain length $R_{max}$, and by its persistence length $\ell_p$, which is the decay length of its orientational memory. The end-to-end distance of a WLC is given by: 
\begin{equation}
<R^2>=2R_{max}\ell_p-2\ell_p^2(1-e^{-R_{max}/\ell_p}).
\label{eq:WLC}
\end{equation}
In order to relate the WLC model to a transport process we use the following relations \cite{rubinstein2003polymer}:
$
R_{max}=\ell N=\ell\frac{t}{t_o}$ and
$\ell_p=-\frac{\ell}{ln(cos(\theta))}\sim\ell\frac{2}{\theta^2}
$, where $N$ is the number of segments in the chain or collisions with obstacles, $\ell$ is the length of a segment or the average distance traveled between collisions, $t_o$ is the average time interval between collisions, and $t$ is the duration of the experiment. The relation between $\ell_p$ and $\theta$, the average change in angle after a collision, is estimated according to the freely rotating chain model \cite{rubinstein2003polymer}. The length and duration of motion in between collisions depends on density. 

At low obstacle densities $n^*<<1$ we may treat the rod as a point particle. The distance between collisions is given by its mean free path, which for uniformly distributed obstacles is $\ell\sim1/n^*$. The duration between collisions is given by $t_o=\ell/v$, where $v$ is the average rod velocity, hence $t_o\sim1/n^*$. In this limit there is no constraint on the scattering angle and the WLC description amounts to a Gaussian chain with $<R^2>=\ell^2N\sim t/n^*$ and $D_{cm}\sim1/n^*$. This is in accord with Enskog theory and simulation results \cite{Hofling2008,Tucker2010}.  

At the limit of infinite obstacle density, assuming infinitely thin rods and point-like obstacles, the rods motion is confined to a straight thin tube. Along that tube the rod barely collides, propagating at approximately constant velocity $\pm v$. The motion of the rod is thus ballistic with $<R^2>=(vt)^2$. This prediction differs from previous predictions \cite{Frenkel1983,Hofling2008,Tucker2010}, showing diffusive motion with $D_{cm}\simeq\sqrt{n^*}$. 

\begin{figure}[h!]
\includegraphics[scale=0.15]{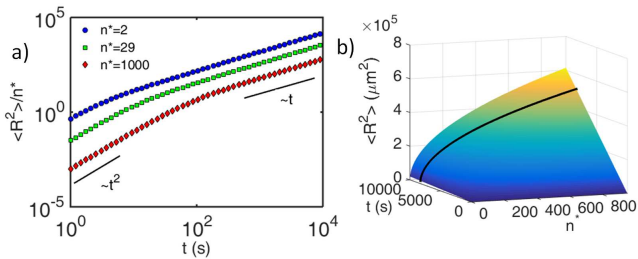}
	\caption{Mean square displacement of a rod moving ballistically in an obstacle field of our WLC inspired model (Eq.~(\ref{eq:WLC_new})). a) Transition from anomalous diffusion to regular diffusion at different $n*$.  b) Mean square displacement as a function of $t$ and $n^*$. At large $t$ the scaling $D_{cm}\sim \sqrt{n^*}$ is recovered (see black solid line). }
	\label{fig:WLC}
\end{figure}

At intermediate obstacle densities the length of the rod becomes larger than the mean free path, and the trajectory shape resembles a WLC configuration. We assume now that $\ell$ is related to the distance traveled parallel to the rod long axis between collision that happen due to perpendicular (or rotational) motion, i.e. $\ell\sim\epsilon$. The scattering angle is constrained and can be estimated by $\theta\simeq\epsilon/L\sim1/\sqrt{n^*}$. The time between collisions becomes $t_o\sim1/\sqrt{n^*}$, the persistence length becomes $\ell_p\sim\sqrt{n^*}$, and $R_{max}\sim t$. Substituting these relations into Eq.~(\ref{eq:WLC}) we have:
\begin{equation}
 \frac{<R^2>}{n^*}\sim 2\frac{t}{\sqrt{n^*}}-2(1-e^{\frac{t}{\sqrt{n^*}}})
 \label{eq:WLC_new}
\end{equation}     
which tailors the two discussed limits of motion: along a straight line, and along a Gaussian chain. A closer look at Eq.~(\ref{eq:WLC_new}) suggests two regimes of motion (Fig.~\ref{fig:WLC}a); for $\frac{t}{\sqrt{n^*}}~<<~0.5$  the motion is super-diffusive, while for $\frac{t}{\sqrt{n^*}}~>>~0.5$ the rod diffuses normally, i.e. the multiple collisions result in a random walk, even though the underlying motion is ballistic. If we take the limit of large $t$ while $n^*$ is kept constant we recover previous results \cite{Frenkel1983,Hofling2008,Tucker2010}, $D_{cm}\sim\sqrt{n^*}$ (Fig.~\ref{eq:WLC}b).  

The superdiffusive regime of motion at short times at intermediate densities identified above (Eq~\ref{eq:WLC_new} and Fig.~\ref{fig:WLC}), relates to the super-diffusive motion we observe for the driven rods (Fig.~\ref{fig:flow}b,c). Here, driving causes the shape of the center-of-mass trajectory to resemble that of a WLC at intermediate timescales.

Following the same lines we can model the motion of rods diffusing in between collisions. At low densities the mean free path scales as $1/n^*$ as before.  $t_o=\sqrt{\ell}/4D_{cm}\sim 1/n^{*2}$, which corresponds to $<R^2>\sim D_{cm} t$ in accord with experiment (Fig.~\ref{fig:Di}b) and simulation \cite{franosch2009}. At large densities the rod is diffusing in a quasi-1D tube, also independent of obstacle density as confirmed by experiment (Fig.~\ref{fig:Di}d) and simulation \cite{franosch2009}. Applying this analysis at intermediate densities results in an unphysical solution (i.e.,  $D_{cm}\sim n^*$). This is expected, since the trajectory shape of diffusing rods differs significantly from WLC configurations, as discussed above. 

In this letter we presented two different experimental realizations of the motion of a rod in a 2D static obstacle field on the single particle level. We later on compare the this motion to the motion of a diffusing rod which is also driven by flow through the obstacle field. Our results agree qualitatively with simulations of diffusing rods \cite{franosch2009}, highlighting two significant differences between theory and experiment. Specifically,  at high density $D_{cm}$ saturates to a smaller value than expected, and $D_\perp$ and $D_r$ do not decay according to the same expected power law. These differences may arise from hydrodynamic interactions, which were not taken into consideration previously, or from the finite size of our rods and obstacles. Our two experimental realizations allow us to characterize such hydrodynamic effects. For example, hydrodynamic interactions near a wall affect the rotational diffusion of a rod much more than the translational diffusion. Another  consequence of hydrodynamic interactions between the rod and the polymer obstacles is that $D_{\parallel}$ decreases with increasing obstacle density even at low densities. In addition, the ratio between parallel and perpendicular diffusion changes gradually with obstacle density for real obstacles. This is in contrast to the more intuitive result obtained for optical obstacles, where the ratio deviates from that of a free rod  only at the onset of confinement $n^*>8$. It should be noted that there is another important difference between both experimental systems that may affect our results, which is the softness of the repulsive potential of the optical scatterers as compared to the hard core repulsion of the polymer pillars. 

By addressing separately the trajectory shape and the transport mechanism we were able to identify a new regime of motion exhibiting super-diffusion for systems with underlying ballistic motion. In addition, the analysis allowed us to pinpoint the significant implications of the different underlying transport mechanisms, both in trajectory shape and area coverage. A better framework to study the transport of a diffusive rod in such systems might be in terms of motion in a percolating cluster or in porous media. Another open question is whether motion on a preassigned trajectory, as assumed in our analysis, is inherently different from motion on a freely chosen trajectory. We note that the use of the WLC description to tailor the transport in high and low obstacle densities is one of many possible choices expressing the orientational memory of the rod's trajectory.      

Finally, the WLC shape of the trajectory of a rod with drift velocity and its apparent superdiffusive motion suggest a connection between a ballistic moving rod and a diffusive one in the presence of driving, at least at short time scales. This may imply that driving (either externally or internally) can lead to enhanced transport of elongated objects in crowded environments. We therefore expect particle shape and especially its aspect ratio to have important implications from the point of view of clogging.

\begin{acknowledgments}
	We thank David Andelman, Haim Diamant and Rigoberto Hernandez for helpful
	discussions. 
\end{acknowledgments}

\end{document}